\documentstyle[preprint,epsfig,prl,aps]{revtex}
\begin{document}
\draft

\title{Photonic Approach to Making a Left-Handed Material }
\author{Gennady Shvets}
\address{Illinois Institute of Technology, Chicago, IL 60616, and Fermi
National Accelerator Laboratory, Batavia, IL 60520}

\newcommand{\ba}{\begin{eqnarray}}
\newcommand{\ea}{\end{eqnarray}}
\newcommand{\be}{\begin{equation}}
\newcommand{\ee}{\end{equation}}
\newcommand{\para}{\parallel}

\maketitle
\begin{abstract}
A new approach to producing a composite material with negative
refraction index is demonstrated. It is shown that a photonic
structure consisting of two dielectric materials, with positive
and negative dielectric permittivities, can support
electromagnetic surface waves which exhibit the unusual
electromagnetic property of left-handedness (or negative
refraction index). Depending on the dielectric materials, these
surface waves localized at the dielectric interfaces can be either
surface plasmons, or phonons. The detailed geometry of the
structure determines whether this composite left-handed material
is isotropic, or anisotropic.
\end{abstract}

\section{Introduction}
\label{sec:intro}

Electromagnetic (EM) properties of materials can be characterized
by two macroscopic quantities: dielectric permittivity $\epsilon$
and magnetic permeability $\mu$. Propagation properties of EM
waves in the material is determined by $\epsilon$ and $\mu$ which
regulate the relationship between the electric field $\vec{E}$ and
the magnetic field $\vec{B}$. Those propagation properties may
depend on the frequency of the waves since both $\epsilon(\omega)$
and $\mu(\omega)$ are, in general, frequency-dependent.

For the overwhelming majority of materials both $\epsilon$ and
$\mu$ are positive in the propagation frequency band. Therefore,
in most materials, as well as in vacuum, the relationship between
$\vec{E}$, $\vec{H}$, and propagation wave vector $\vec{k}$ is
given by the right hand rule: $\vec{k} \cdot [\vec{E} \times
\vec{H}] > 0$. The consequence of this is that the group velocity
$\vec{v}_g = \vec{P}/U$ (where $\vec{P} = c [\vec{E} \times
\vec{H}]/4\pi$ is the Poynting flux of the wave and $U$ is the
wave energy density) of the propagating wave packet points in the
same direction as its phase velocity $\vec{v}_{\rm ph} = \omega
\vec{k}/|\vec{k}|^2$.

It was first pointed out by Veselago~\cite{veselago} that wave
propagation is also possible in materials which have
simultaneously negative $\mu$ and $\epsilon$. Since in such
environments the relationship between $\vec{E}$, $\vec{H}$, and
$\vec{k}$ is given by the left-hand rule, such materials are
referred to as the Left-Handed Materials (LHMs). Their
electromagnetic properties are significantly different from those
of the right-handed materials because the group and phase
velocities of electromagnetic waves in LHMs oppose each other:
$\vec{k} \cdot \partial \omega/\partial \vec{k} < 0$. For example,
an electromagnetic wave incident on an interface between the right
and left-handed materials stays on the same side of the interface
normal~\cite{veselago}. In other words, its refracted angle is
{\it negative}. This property gives rise to another name for the
LHMs: materials with negative index of
refraction~\cite{smith_kroll}. Also, the sign of the Doppler
effect is reversed in LHMs: an approaching source appears to emit
higher frequency waves than the receding one.

LHMs have recently attracted significant attention because of
their promise for developing the so called ``perfect''
lenses~\cite{pendry_lens} and low reflectance
surfaces~\cite{smith_kroll}. ``Perfect'' lens enables focusing
electromagnetic waves to a spot size much smaller than the
wavelength $\lambda$. Conventional lenses do not permit the image
to be significantly sharper than $\lambda/2$.

Since materials with $\mu < 0$ do not naturally occur, LHM has to
be artificially constructed. An LHM in the microwave frequency
band was recently constructed~\cite{smith_exper} as an array
consisting of metal rods (which provided $\epsilon <
0$~\cite{rothman}) and split-ring resonators (which provided $\mu
< 0$~\cite{pendry_magnet}). Photonic structures have also been
known to exhibit negative group velocity due to the band folding
effect~\cite{notomi}. Below I introduce a new concept of making an
LHM by creating a photonic structure consisting of dielectric
regions with $\epsilon < 0$ separated by thin vacuum gaps.
Left-handedness of such structures is due to the existence of the
surface waves at the vacuum/dielectric interfaces. Depending on
the nature of the negative $\epsilon$ material used, those surface
waves could be either surface plasmons, or surface phonons.

Practical implementation of such photonic structures at a micron
scale is encouraged by the availability of low-loss dielectrics
and semiconductors with negative $\epsilon$, including many polar
crystals such as SiC, LiTaO$_3$, LiF, and ZnSe. The
frequency-dependent dielectric permittivity of these crystals,
given by the approximate formula~\cite{kittel} $ \epsilon(\omega)
= \epsilon_{\infty} (\omega^2 - \omega_L^2)/(\omega^2 -
\omega_T^2),$ is negative for $\omega_T < \omega < \omega_L$.
Another example of a medium with $\epsilon < 0$ is the free
electron gas with $\epsilon = 1 - \omega_p^2/\omega^2$, where
$\omega_p$ is the plasma frequency. Its dielectric permittivity
turns negative for $\omega < \omega_p$.

The major innovation introduced in this paper is the use of the
dielectric materials with small (of order $-1$) negative
dielectric permittivity in order to construct a left-handed
composite material. This is significantly different from earlier
work~\cite{smith_exper} where a periodic arrangement of metallic
components was assembled to achieve left-handedness in the
microwave frequency range. For microwave, as well as infrared,
frequencies the dielectric permittivity of metals is essentially
$\epsilon = - \infty$.

\begin{figure}[h]
\centering {\hspace{-0pt}
\epsfig{file=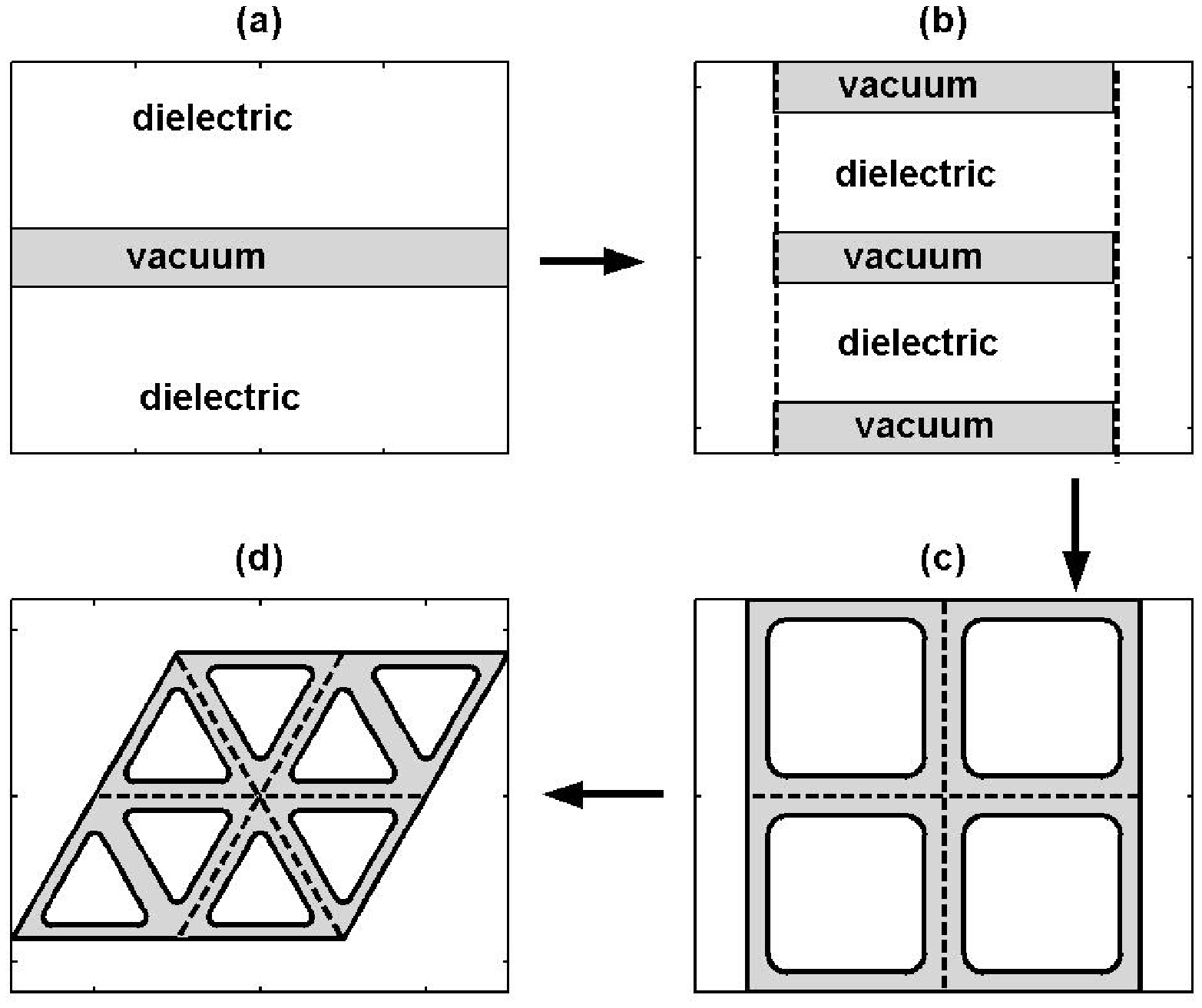,angle=-0}}
\vspace{10pt}
\caption{(a) Single dielectric waveguide (SDW)
consists of a vacuum gap surrounded by the dielectric cladding with
$\epsilon_c < 0$. (b) One-dimensional waveguide stack (WS). (c)
Square lattice photonic waveguide (SLPW): dielectric cladding
regions (white squares) separated by vacuum channels (shaded). (d)
Triangular lattice photonic waveguide (TLPW). In (a)-(d): white
regions represent dielectric cladding, shaded regions represent
vacuum channels.}
\label{fig:figure_structures}
\end{figure}

\begin{figure}[h]
\centering {\hspace{-0pt} \epsfig{file=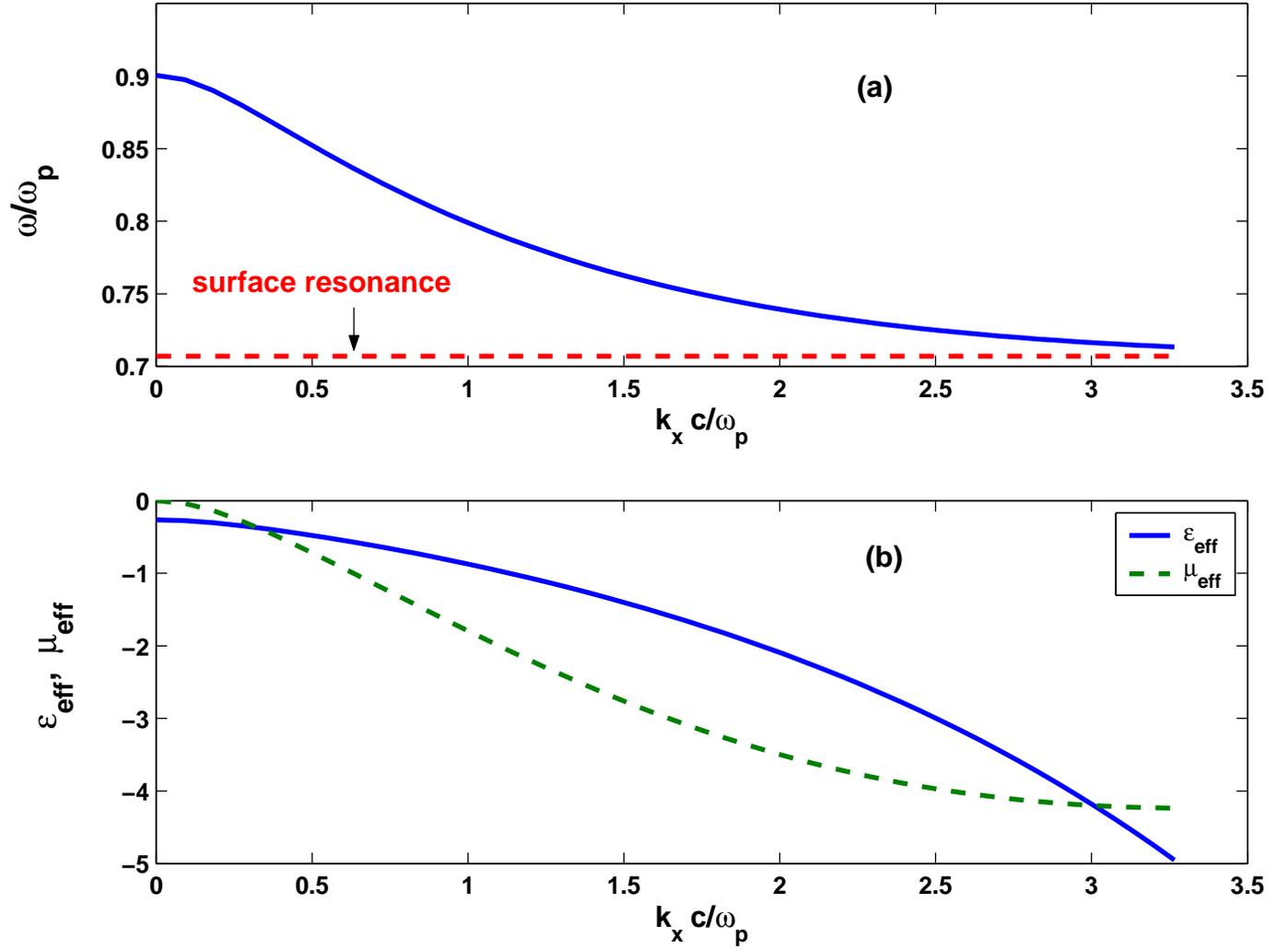,angle=-0}}
\vspace{10pt} \caption{(a) Dispersion relation, and (b) effective
dielectric permittivity $\epsilon_{\rm eff}$ and magnetic
permeability $\mu_{\rm eff}$. Dielectric permittivity of cladding
$\epsilon_c = 1 - \omega_p^2/\omega^2$, gap width $2b =
c/\omega_p$.} \label{fig:1dchannel}
\end{figure}

The organization of the remainder of the paper is schematically
shown in Fig.~\ref{fig:figure_structures}. In
Section~\ref{sec:sdw} a single dielectric waveguide (SDW) with
$\epsilon_c < 0$ dielectric cladding
(Fig.~\ref{fig:figure_structures}(a)), is shown to exhibit
left-handedness due to the existence of the surface waves at the
vacuum-cladding interface. The effective $\epsilon_{\rm eff}$ and
$\mu_{\rm eff}$ of the waveguide are shown to be negative (see
Fig.~\ref{fig:1dchannel}). Two types of surface waves are
considered: surface plasmons at the vacuum/plasma interface, and
surface phonons at the interface between vacuum and silicon
carbide (SiC). Due to its remarkable property of having the
dielectric permittivity $\epsilon = -1$ at the wavelength ($10.6
\mu$m) produced by the conventional CO$_2$ lasers, silicon carbide
is also proposed as the solid state medium for making a near-field
``perfect'' lens: the tool for enhanced near-field imaging in
mid-infrared (see Fig.~\ref{fig:perf_lens1}).

\begin{figure}[h]
\centering {\hspace{-0pt} \epsfig{file=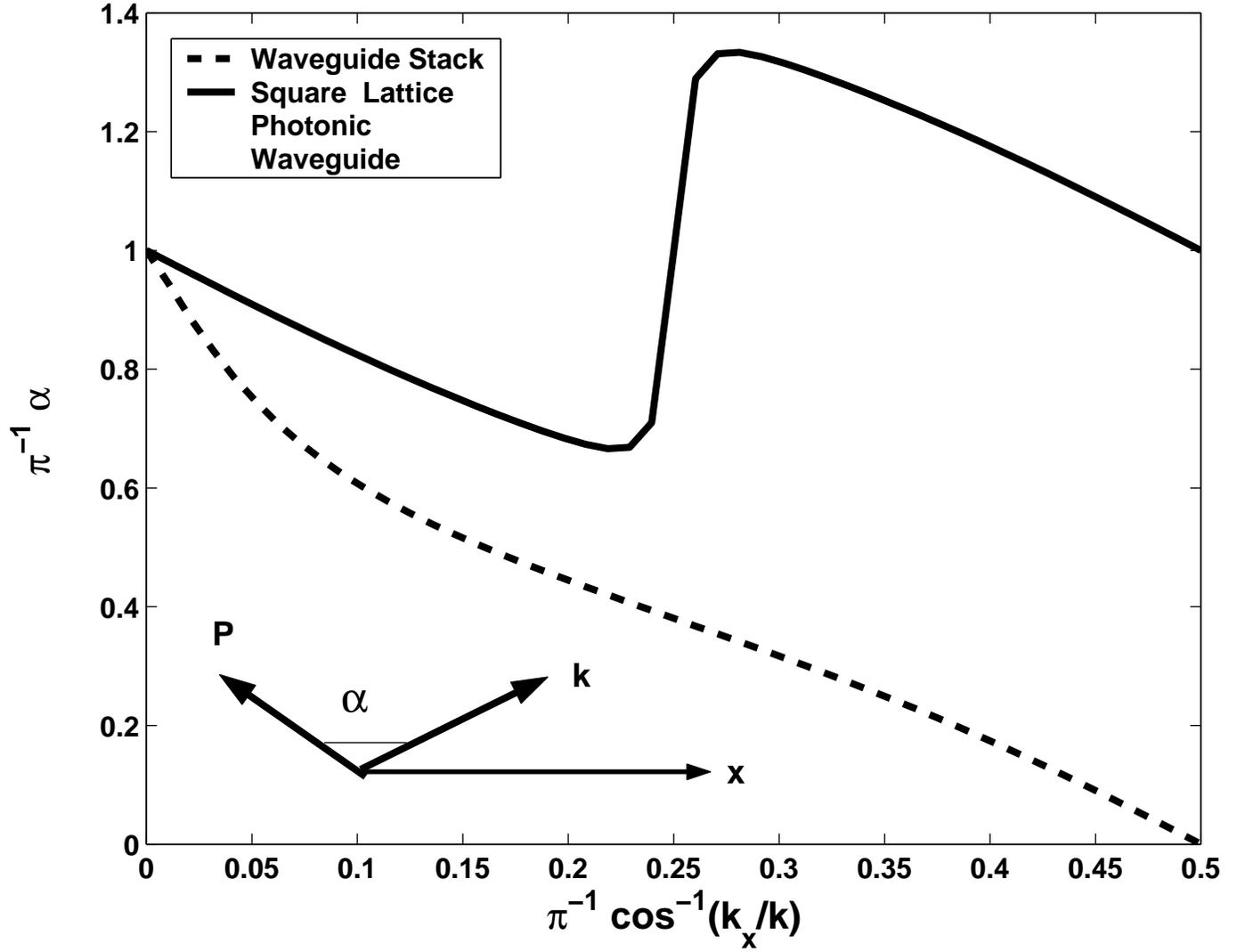,angle=-0}}
\vspace{10pt} \caption{Angle between the phase and group
velocities $\alpha$ v.~s. the propagation angle $\theta =
\cos^{-1}{k_x/k}$, where $k = d^{-1} \pi/4$. WS parameters:
$\omega_p d/c =4$, $b/d = 1/8$. SLPW parameters: $\omega_p d =
4.8$, $b/d = 1/8$, $r_b = b$.} \label{fig:poynt_angle}
\end{figure}

Of course, surface waves can only propagate along (and not across)
the waveguide walls. Section~\ref{sec:ws} consider a stack of
dielectric waveguides (Fig.~\ref{fig:figure_structures}(b)), which
supports waves propagating in all directions. Only the waves
propagating in a limited range of directions, however, exhibit the
$\vec{v}_g \cdot \vec{v}_{\rm ph} < 0$ property (where $v_g$ and
$v_{\rm ph}$ are the group and phase velocities). Therefore, the
waveguide stack (WS) is not a proper LHM, and the two-dimensional
photonic waveguides shown in Fig.~\ref{fig:figure_structures}(c,d)
are considered in Sections~\ref{sec:slpw},\ref{sec:tlpw} as the
further refinements of the concept. The square lattice photonic
waveguide (SLPW) turns out to be a highly anisotropic LHM, with
the angle $\alpha = \angle (\vec{v}_{\rm ph},\vec{v}_{\rm g})$
always satisfying $\pi/2 < \alpha < 3 \pi/2$ but strongly
dependent on the propagation direction (see
Fig.~\ref{fig:poynt_angle}). SLPW is anisotropic even for small
wavenumbers. This result is explained using the standard
perturbation theory. The triangular lattice photonic waveguide
(TLPW) is found to be a perfectly isotropic LHM for small
wavenumbers $k_{x,y} \ll \pi/d$, where $d$ is the lattice
periodicity. Recently~\cite{valanju}, it has been suggested that
the group velocity in LHMs is not aligned with the phase velocity,
making the perfect lens impossible. It is constructively
demonstrated that this is not the case even for an artificially
constructed LHM shown in Fig.~\ref{fig:figure_structures}(d). The
main results are summarized in Sec.~\ref{sec:concl}.

\section{Propagation of Surface Plasmons and Phonons in a Single
Dielectric Waveguide}
\label{sec:sdw}

To illustrate how a negative $\mu$ can be mimicked using only
dielectrics with negative $\epsilon$, consider electromagnetic
wave propagation in the horizontal ($x$) direction in a dielectric
waveguide shown in Fig.~\ref{fig:figure_structures}(a), with a
piecewise constant dielectric constant: $\epsilon = 1$ in the
vacuum channel (for $-b < y < b$) and $\epsilon = \epsilon_c < 0$
inside the cladding (for $|y| > b$). Consider a confined
transverse magnetic (TM) wave with non-vanishing components
($E_x$, $E_y$, $H_z$), and assume, by symmetry, that $E_y = H_z =
\partial_y E_x = 0$ at $y = 0$. Since we are interested in the
wave propagation along $x$, introduce integrated over the
transverse direction $y$ quantities
\[
\tilde{E}_{x,y} = \int_0^{\infty} dy \ E_{x,y}, \ \ \ \
\tilde{H}_{z} = \int_0^{\infty} dy \ H_{z}.
\]
From Faraday's and Ampere's laws, assuming that $E_x(x \rightarrow
\infty) = 0$ and integrating by parts, obtain, correspondingly,
\be
\frac{\partial \tilde{E}_y}{\partial x} =
\frac{i \omega}{c} \mu_{\rm eff} \tilde{H}_z, \ \ \ \
\frac{\partial
\tilde{H}_z }{\partial x} = \frac{i \omega }{c}
\epsilon_{\rm eff} \tilde{E}_y,
\label{eq:ampere1}
\ee
where the effective dielectric permittivity and magnetic permeability
of a dielectric waveguide are defined as follows:
\be
\epsilon_{\rm eff} =
\frac{1}{\tilde{E}_y} \int_0^{\infty} dy \epsilon E_y, \ \ \
\mu_{\rm eff} = 1 + \frac{i c}{\omega}
\frac{E_x(y=0)}{\tilde{H}_z}.
\label{eq:epsilon_eff}
\ee
The definition of the weight-averaged $\epsilon$ is intuitive, and
$\mu_{\rm eff}$ is defined so as to eliminate the longitudinal component of
the electric field $E_x$ which does not contribute to the power flow along
the waveguide.

Equations (\ref{eq:ampere1}) yield $k_x^2 c^2/\omega^2 = \mu_{\rm
eff} \epsilon_{\rm eff}$, necessitating that $\epsilon_{\rm eff}$
and $\mu_{\rm eff}$ be of the same sign for a propagating wave.
Note that $\mu_{\rm eff}$ is a complicated function of the channel
width, frequency, and the dielectric constant of the cladding, and
could generally be either positive or negative. Also, the
propagating mode is left-handed only if $\mu_{\rm eff} < 0,
\epsilon_{\rm eff} < 0$, necessitating that the dielectric
constant of the cladding be negative.

Calculating $\epsilon_{\rm eff}$ and $\mu_{\rm eff}$ requires the
exact mode structure in the waveguide. The equation for $H_z$ is
\begin{equation}
\frac{\partial}{\partial x} \left( \frac{1}{\epsilon}
\frac{\partial H_z}{\partial x} \right) +
\frac{\partial}{\partial y} \left( \frac{1}{\epsilon}
\frac{\partial H_z}{\partial y} \right)
= - \frac{\omega^2}{c^2}  H_z,
\label{eq:by1}
\end{equation}
and for the SDW case the harmonic dependence $H_z \propto
\exp{i(k_x x - \omega t)}$ is assumed. Introducing $\chi_v =
\sqrt{\omega^2/c^2 - k_x^2}$, $\chi_p = \sqrt{k_x^2 - \epsilon_c
\omega^2/c^2}$ for $k_x c < \omega$, the mode structure is given
by $H_z = \sin{\chi_v y}/\sin{\chi_v b}$ for $0 < y < b $ and $H_z
= \exp{[-\chi_p(y-b)]}$ for $b < y < \infty $. It is a surface
wave because it is localized near the vacuum-cladding interface.
Continuity of $E_x$ requires the continuity of $\epsilon^{-1}
\partial_y H_z$ across the vacuum-cladding interface at $y = b$.
The dispersion relation $\omega$ v.~s. $k_x$ is found by solving
the boundary condition equation $\chi_p = -\epsilon_c
\chi_v/\tan{\chi_v b}$ simultaneously with the equations for
$\chi_v$ and $\chi_p$. Expressions for the effective permittivity
and permeability are given by
\ba
&&\epsilon_{\rm eff} = \left(
\frac{1}{\chi_p} + \frac{1 - \cos{\chi_v b}}{\chi_v \sin{\chi_v
b}} \right) \left( \frac{1}{\epsilon_c \chi_p} + \frac{1 -
\cos{\chi_v b}}{\chi_v
\sin{\chi_v b}} \right)^{-1} \nonumber \\
&&\mu_{\rm eff} = 1 - \frac{c^2 \chi_x}{\omega^2 \sin{\chi_v b}}
\left( \frac{1}{\epsilon_c \chi_p} + \frac{1 - \cos{\chi_v
b}}{\chi_v \sin{\chi_v b}} \right)^{-1}, \label{eq:mu_eff2} \ea
and similar equations are obtained for the sub-luminal mode with
$k_x c > \omega$.

Two cladding materials are examined below: a plasmonic material,
and a polar dielectric silicon carbide (SiC). Plasmonic materials
are characterized by the frequency-dependent dielectric
permittivity $\epsilon_c = 1 - \omega_p^2/\omega (\omega + i
\gamma)$, where $\omega_p$ is the plasma frequency, and $\gamma$
is the damping constant. At optical frequencies most metals can be
considered plasmonic materials. Surface plasmons exist at the
interface between a plasmonic material and another dielectric (or
vacuum) with positive dielectric permittivity. Silicon carbide is
a low-loss polar dielectric which exhibits Restsrahlen: it's
dielectric permittivity is negative for frequencies $\omega_T <
\omega < \omega_L$, where $\omega_T = 793$cm$^{-1}$ and $\omega_L
= 969$cm$^{-1}$.

\subsection{Plasmonic Dielectric Cladding}
\label{sec:plasmonic}

The dispersion relation and the corresponding $\epsilon_{\rm eff}$
and $\mu_{\rm eff}$ are plotted in Fig.~\ref{fig:1dchannel}(a,b)
for a SDW with plasma-like cladding ($\epsilon_c(\omega) = 1 -
\omega_p^2/\omega^2$) and the gap width $2b = c/\omega_p$. The
propagating surface mode in a SDW exists at the vacuum/plasma
interface -- therefore it is a surface plasmon. This surface
plasmon exhibits left-handedness: it's group velocity $v_g =
\partial \omega/\partial k < 0$ is negative, and so are $\mu_{\rm
eff} < 0$ and $\epsilon_{\rm eff} < 0$. The cutoff at $\omega =
0.9 \omega_p$ is caused by the vanishing of the $\mu_{\rm eff}$.
This is very different from the cutoff at $\omega = \omega_p$ in a
homogeneous plasma-like medium which is caused by the vanishing of
$\epsilon_{\rm eff}$. Note that {\it two} vacuum/plasma interfaces
are required to make a surface plasmon left-handed.

Why is $v_g < 0$ despite $v_{\rm ph} = \omega/k > 0$? The total
Poynting flux $P_x = c E_y H_z/4\pi$ along the dielectric
waveguide is the sum of the fluxes inside the cladding and in the
vacuum gap. In the gap, $E_y$ and $H_z$ are in phase, so $P_x >
0$. Inside the cladding $E_y$ reverses its sign across the
vacuum/cladding interface because $\epsilon_c < 0$, and the
electric displacement vector $D_y = \epsilon E_y$ is continuous.
$H_z$ is continuous across the interface, resulting in $P_x < 0$
in the cladding. For a narrow gap, the integrated Poynting flux is
negative, and the mode is left-handed.

We emphasize that not any surface wave is left-handed. Achieving
left-handedness requires that (a) the frequency of the mode lies
within the stop-band of the cladding (so that $\epsilon_c < 0$),
(b) there are two interfaces, and (c) the vacuum gap between the
interfaces is small. The gap between the two claddings with
$\epsilon_c < 0$ need not be vacuum -- it can be filled with
another dielectric which has a positive dielectric permittivity to
enable the existence of the surface waves. The cladding with
$\epsilon_c < 0$ also needs not be plasma-like. In the next
section the silicon carbide cladding is considered.

\subsection{Silicon Carbide Dielectric Cladding}
\label{sec:sic}

Silicon carbide (SiC) is a polar dielectric~\cite{spitzer_sic}
with very interesting photonic properties because its dielectric
permittivity is of order $\epsilon \sim -1$ for the wavelengths
close to $10$ microns. Therefore, a micron-scale waveguide with
SiC cladding supporting left-handed surface phonons with the
frequency corresponding to the vacuum wavelength of $10 \mu$m can
be envisioned. The frequency dependent dielectric permittivity of
SiC is given by~\cite{spitzer_sic}
\begin{equation}
\epsilon_c  = \epsilon_{\infty} \frac{\omega_L^2 - \omega^2 + i
\gamma \omega}{\omega_T^2 - \omega^2 + i \gamma \omega},
\label{eq:eps_sic}
\end{equation}
where $\epsilon_{\infty} = 6.7$, $\omega_L = 969$cm$^{-1}$,
$\omega_T = 793$cm$^{-1}$, and $\gamma = 4.76$cm$^{-1}$.

\begin{figure}[h]
\centering {\hspace{-0pt} \epsfig{file=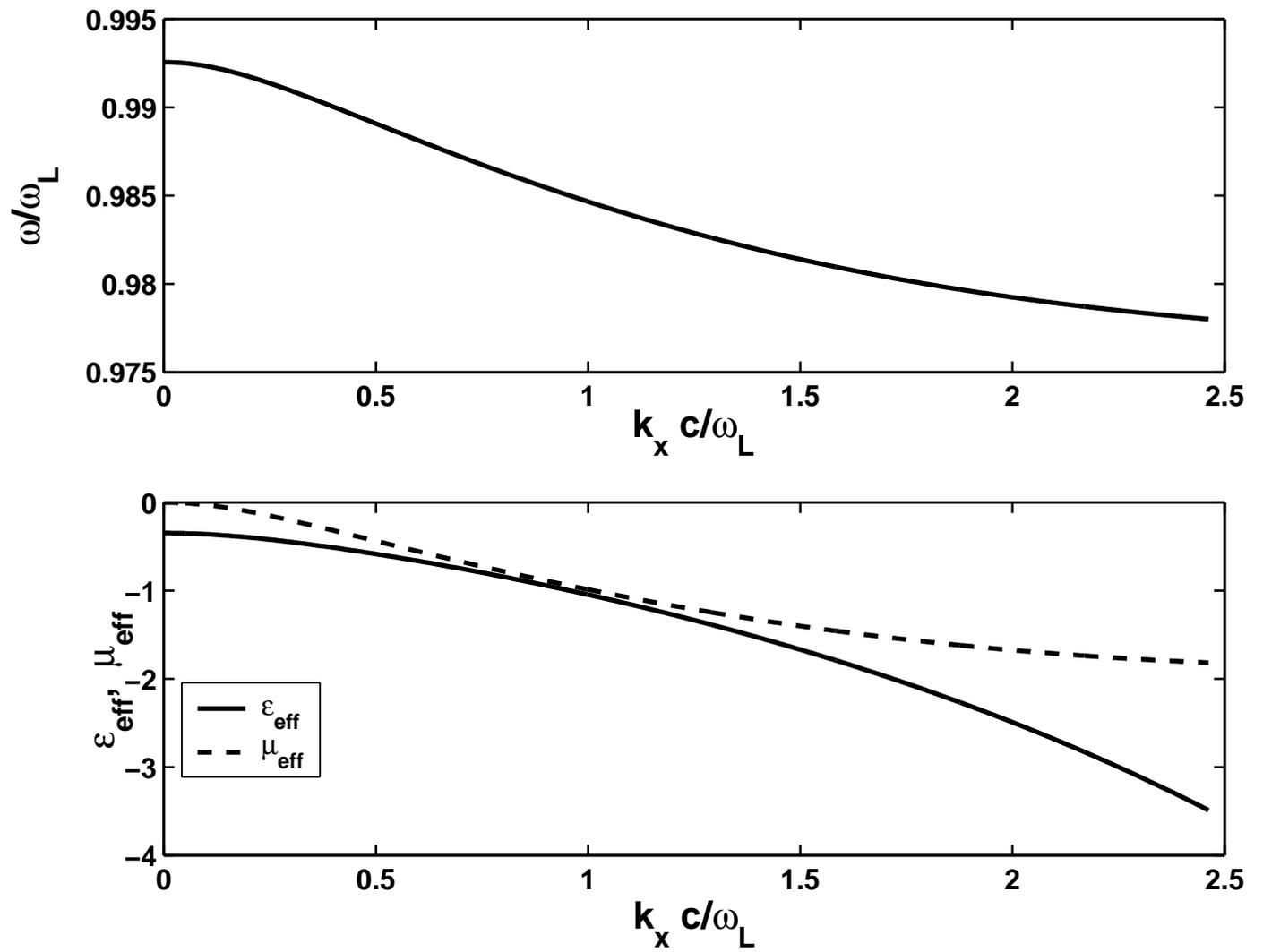,angle=-0}}
\vspace{10pt} \caption{(a) Dispersion relation, and (b) effective
dielectric permittivity $\epsilon_{\rm eff}$ and magnetic
permeability $\mu_{\rm eff}$ of a dielectric waveguide with SiC
cladding. Gap width $2b = c/\omega_L$.} \label{fig:sic_channel}
\end{figure}

The dispersion relation and the corresponding $\epsilon_{\rm eff}$
and $\mu_{\rm eff}$ are plotted in Fig.~\ref{fig:sic_channel}(a,b)
for a SDW with the SiC cladding and the gap width $2b =
c/\omega_L$. The small damping constant $\gamma \approx 0.005
\omega_L$ was neglected. The left-handed surface mode is a surface
phonon. Qualitatively, there is a similarity between the
electromagnetic properties of a left-handed plasmon (shown in
Fig.~\ref{fig:1dchannel}) and those of a left-handed surface
phonon (shown in Fig.~\ref{fig:sic_channel}). Both exhibit
negative group velocity $v_g = \partial \omega/\partial k < 0$ and
negative $\epsilon_{\rm eff}$, $\mu_{\rm eff}$.

One essential difference is that the group velocity of the surface
phonon is much smaller than that of the surface plasmon. This is
related to the large energy density of the polaritons (coupled
phonon polarization and photon waves) in SiC. The energy density
is proportional to $U \propto \partial (\omega \epsilon)/\partial
\omega$ which is proportional to $\omega_L^2
\epsilon_{\infty}/(\omega_L^2 - \omega_T^2)$. Both the large value
of $\epsilon_{\infty}$ and the narrow width of the reststrahlen
band (between $\omega_T$ and $\omega_L$) contribute to the small
group velocity of the surface phonons.

Note that there exists a frequency for which both $\mu_{\rm eff}$
and $\epsilon_{\rm eff}$ are approximately equal to $(-1)$. A
composite material with $\mu_{\rm eff} = \epsilon_{\rm eff} = -1$
was shown to be ideal for making a ``perfect''
lens~\cite{pendry_lens} which is capable of imaging objects much
smaller than the wavelength of light.

Another type of a near-field lens which does not require $\mu_{\rm
eff} = -1$ (although still requires $\epsilon_{\rm eff} = -1$) was
also recently suggested~\cite{pendry_lens,platzman_lens}. It was
concluded that a thin slab of material with $\epsilon = -1$ can
produce images of sub-wavelength objects with the resolution which
by far exceeds that of the conventional lenses or even near-field
imaging. Unfortunately, finding the appropriate material with
$\epsilon = -1$ is challenging. Gaseous plasma was suggested as a
possible candidate for enhanced near-field imaging in the
microwave frequency range~\cite{platzman_lens}.

Unfortunately, the microwave frequency (or the resolution
accomplished with microwaves) may be too low for many
applications. Another promising idea of using thin metal
layers~\cite{pendry_lens} for enhanced near-field imaging requires
extremely thin (less than $100$ Angstroms) films and very short
wavelengths ($0.15 \mu$m for Ag films). The reason I've chosen SiC
as the exemplary ingredient for making a left-handed material is
that it may be interesting from the technological standpoint: its
dielectric permittivity $\epsilon = -1$ for the frequency
corresponding to the vacuum wavelength of $\lambda_0 \equiv 2\pi
c/\omega_0 = 10.55 \mu$m. This wavelength is produced by the
widely available CO$_2$ lasers. To illustrate the resolution
achievable by the combination of a CO$_2$ laser and a thin film of
SiC, consider the image of a narrow (sub-wavelength) slit produced
by a film of SiC of the width $d = \lambda_0/8\pi = 0.42 \mu$m.

Two-dimensional geometry is assumed, and a thin source at $x=0$ is
assumed to be infinitely extended in the $z-$direction while
having the Gaussian shape in the $y-$ direction. Transverse
magnetic (TM) mode (with field components $H_z$, $E_x$, and $E_y$)
is assumed, and the prescribed magnetic field at the slit $H_z =
\exp{(-y^2/\sigma^2)}$ with $\sigma = \lambda_0/16\pi = 0.21 \mu$m
is assumed. The thin SiC film is located between $x = d/2$ and $x
= 3d/2$.

\begin{figure}[h]
\centering {\hspace{-0pt}
\epsfig{file=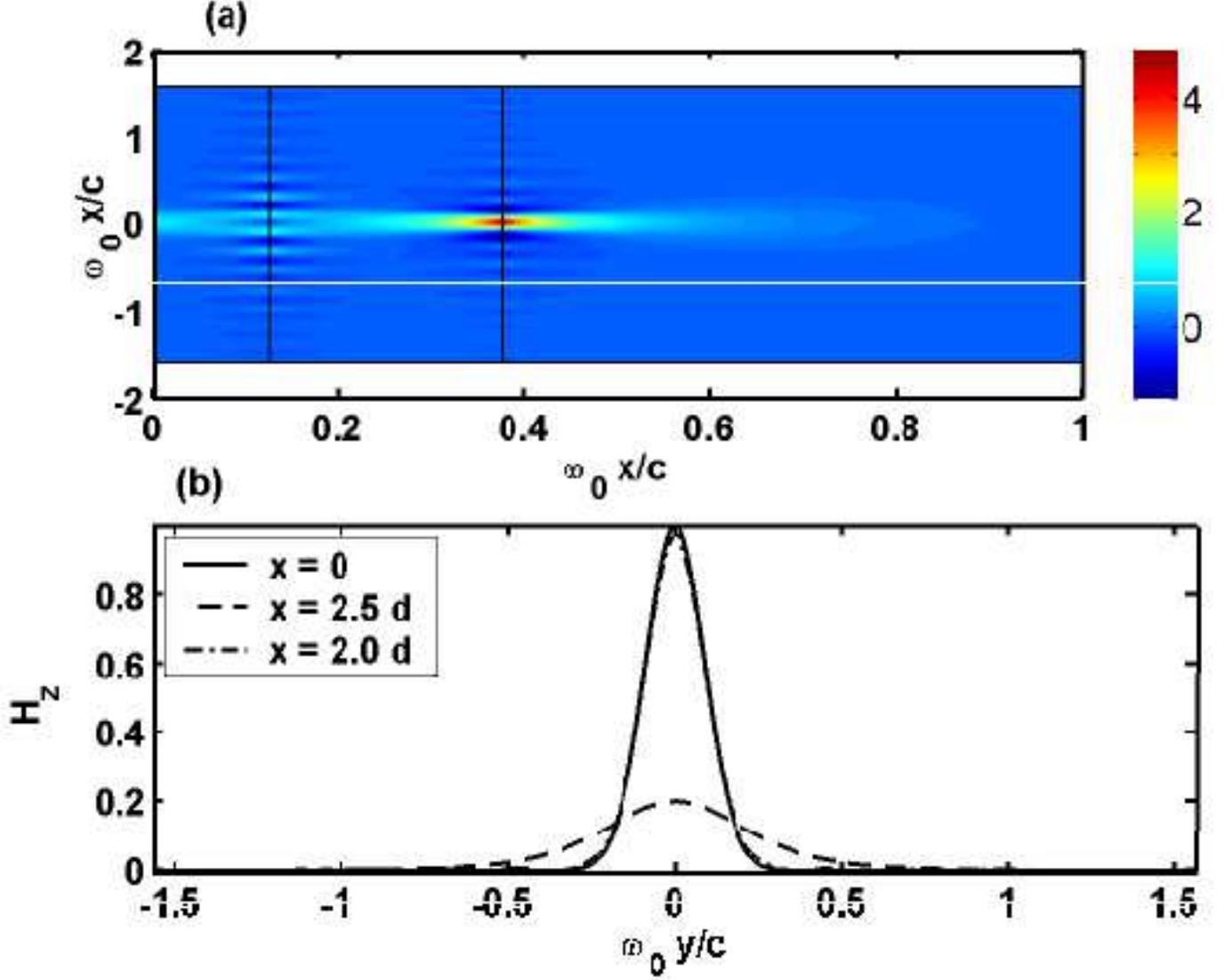,angle=-0}} \vspace{10pt}
\caption{Magnetic field $H_z$ produced by the Gaussian source of
the size $\sigma = 0.125 c/\omega_0$ located in the $x = 0$ plane.
A slab of SiC with $\epsilon = -1$ of width $d = 0.25 c/\omega_0$
is located between $x = d/2$ and $x = 1.5 d$. The image plane is
at $x = 2 d$. (a) Magnetic field $H_z(x,y)$ distribution inside
the simulation domain $0 < x < c/\omega_0$, $-\lambda_0/2 < x <
\lambda_0/2$ (in color online) and (b) lineouts of $H_z(y)$ in
three different planes: $x = 0$ (source plane, solid line), $x =
2d$ (image plane, dot-dashed line), and $x = 2.5 d$ (dashed
line).} \label{fig:perf_lens1}
\end{figure}

The two-dimensional distribution of the magnetic field $H_z(x,y)$
in the rectangular area $0 < x < \lambda_0/2\pi$ and $-\lambda_0/2
< y < \lambda_0/2$ is shown in Fig.~\ref{fig:perf_lens1}(a). The
profiles of $H_z(y)$ in three planes: at $x=0$ (which represents
the original shape of the sub-wavelength slit), at $x=2 d$ (the
imaging plane), and $x = 3d$ (the SiC slab width behind the image
plane), are plotted in Fig.~\ref{fig:perf_lens1}(b). The enhanced
near-field image at $x = 2d$ is practically indistinguishable from
the shape of the slit at $x=0$, and is of much higher fidelity
than the un-enhanced near-field at $x = 3d$. Therefore, this
numerical example demonstrates the higher fidelity of the enhanced
near-field imaging in comparison with the standard one.

To my knowledge, the idea of using a low-loss polar dielectric
with the reststrahlen band for enhanced near-field infrared
imaging with sub-micron resolution is presented here for the first
time. Possible applications include nano-lithography using a
CO$_2$ laser and high-resolution biological and chemical sensors.

\section{Waveguide Stack}
\label{sec:ws}

The SDW was considered merely to elucidate the emergence of the
left-handed surface waves, and to derive the necessary conditions
(a-c) for their existence. The objective of this Letter is to
construct a photonic structure capable of transmitting infinitely
extended (planar) left-handed waves in all directions. SDW is not
infinitely extended in $y-$ direction, and it only enables wave
propagation in $x-$direction. Thus, the waveguide stack (WS) with
periodicity $d$ in $y-$direction, shown in
Fig.~\ref{fig:figure_structures}(b), is considered next. The
elementary cell of a WS is assumed to have the piecewise constant
dielectric permittivity: $\epsilon(x) = 1$ for $|x| < b$ and
$\epsilon(x) = \epsilon_c$ for $b < |x| < d/2$. Electromagnetic
waves in a WS are characterized by the two numbers: phase shift
per unit cell $-\pi < \phi_y \equiv k_y d < \pi$ and the
wavenumber $\vec{k}_x$. Electromagnetic waves in a periodically
layered medium have been intensely studied since the
70's~\cite{yeh}, but without the emphasis on surface waves which
can exhibit left-handedness.

Assuming that $\epsilon_c = 1 - \omega_p^2/\omega^2$, $\omega_p
b/c = 0.5$, and $\omega_p d/c = 4$, the dispersion relation in the
WS $\omega \equiv \omega(k_y,k_x)$ was numerically calculated by
integrating Eq.~(\ref{eq:by1}) between $y=0$ and $y=d$ and
requiring that $H_z(y=d)/H_z(y=0) = \exp{i\phi_y}$.  It is found
that $\partial \omega/\partial k_x < 0$, $\partial \omega/\partial
k_y > 0$. Therefore, waves propagate as left-handed along the
$x$-direction, and as right-handed along the $x-$ direction. The
angle $\alpha$ between the phase and group velocities in the WS is
plotted as the dashed line in Fig.~\ref{fig:poynt_angle} v.~s.~the
angle $\theta = \cos^{-1}{k_x/|\vec{k}|}$ between the wavenumber
$\vec{k} = k_x \vec{e}_x + k_y \vec{e}_y$ and the $x-$axis. For
small $\theta$, $\vec{v}_{\rm g} \cdot \vec{v}_{\rm ph} < 0$
($\alpha > \pi/2$), and the waves are essentially left-handed.
However, for large $\theta$, $\alpha < \pi/2$, and the waves are
right-handed. Therefore, the waveguide stack is not an LHM.

\section{Electromagnetic Properties of Square Lattice Photonic Waveguide}
\label{sec:slpw}

To ensure that surface waves can propagate in $y-$ direction, an
additional set of vertical channels is introduced, as shown in
Fig.~\ref{fig:figure_structures}(c). Rectangular regions of the
dielectric are separated from each other by a network of
orthogonal vacuum channels which form a square lattice with the
period $d = 4.8 c/\omega_p$. The channel widths are $b = d/8$, and
the dielectric boundaries are rounded with the radius $r_b = b$.
Wave propagation along the resulting square lattice photonic
waveguide (SLPW) is characterized by two phase shifts per cell,
$\phi_x \equiv k_x d$ and $\phi_y = k_y d$. Equation
(\ref{eq:by1}) is solved for eigenvalues using a finite elements
code FEMLAB~\cite{ref_femlab}. The solution satisfies the
following boundary conditions: $H_z(d/2,y)/H_z(-d/2,y) =
e^{i\phi_x}$ and $H_z(x,d/2)/H_z(x,-d/2) =
e^{i\phi_y}$~\cite{smirnova_jap}. The local $\vec{P}(x,y)$
and cell-averaged $\vec{P}_{\rm av}$ Poynting fluxes are computed
for each solution. Since $\vec{v}_{\rm g} \para \vec{P}_{\rm av}$,
the angle $\alpha = \angle (\vec{k},\vec{P}_{\rm av})$ between
$\vec{P}_{\rm av}$ and $\vec{k} \para \vec{v}_{\rm ph}$
is used to classify waves into left- and
right-handed.

\begin{figure}[h]
\centering {\hspace{-0pt} \epsfig{file=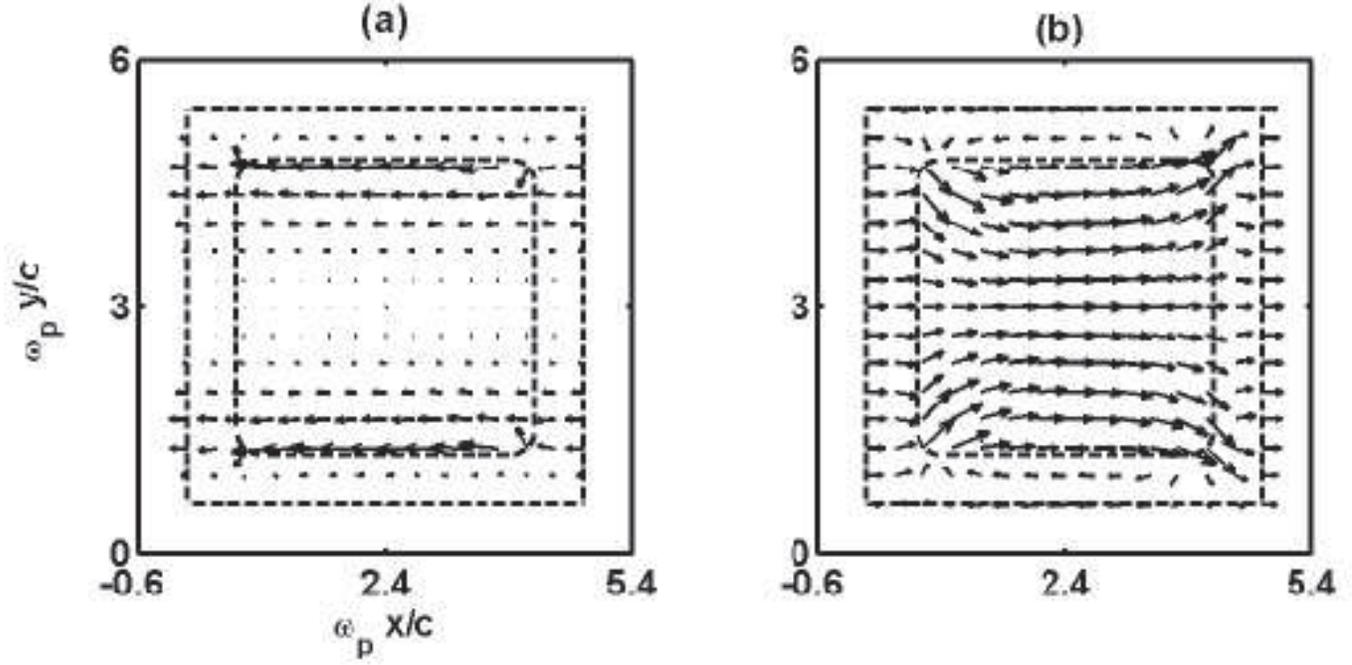,angle=-0}}
\vspace{4pt} \caption{ Poynting flux in SLPW for the modes with
$k_x = d^{-1} \pi/4$, $k_y = 0$: (a) lower-frequency left-handed
mode, $\vec{P}_{\rm av} \cdot \vec{k}  < 0$, $\omega/\omega_p =
0.85$; (b) higher-frequency right-handed mode, $\vec{P}_{\rm av}
\cdot \vec{k}  > 0$, $\omega/\omega_p = 0.95$. Structure
parameters: cell size $d=4.8 c/\omega_p$, channel width $b =
d/4$.} \label{fig:left_right}
\end{figure}

Two classes of electromagnetic modes were found for any given
wavenumber $\vec{k}$. The higher-frequency mode is right-handed,
and the lower-frequency mode is left-handed. The physical
difference between these modes is illustrated by
Fig.~\ref{fig:left_right}, where the local Poynting fluxes are
shown for $k_x = d^{-1} \pi/4$, $k_y = 0$. The Poynting flux of
the left-handed mode ($\omega = 0.85 \omega_p$) shown in
Fig.~\ref{fig:left_right}(a) is localized near the vacuum/cladding
interfaces, and points in the negative direction. The mode clearly
propagates as a surface wave. In contrast, the right-handed mode
($\omega = 0.95 \omega_p$) propagates by tunneling across the
dielectric cladding, which is why the Poynting flux is positive
and not localized, as shown in Fig.~\ref{fig:left_right}(b). The
two modes have exactly the same frequency $\omega(\vec{k} = 0)
\equiv \omega_c = 0.852 \omega_p$ (i.~e.~ are degenerate) for
$\vec{k} = 0$. As will be demonstrated below, the mode degeneracy
results in the strong anisotropy of the propagating modes with
small (but finite) wavenumbers $\vec{k}$: the mode frequency not
only depends on $k = |\vec{k}|$, but also on its direction
$\vec{n} = \vec{k}/k$.

The angle $\alpha = \angle (\vec{k},\vec{P}_{\rm av})$ is plotted
in Fig.~\ref{fig:poynt_angle} as the function of the propagation angle
$\theta$ for the lower-frequency mode of the
SLPW. In an {\it isotropic} LHM $\alpha = \pi$ for any $\theta$.
In general, the medium need not be isotropic, in
which case we require that $\pi/2 < \alpha < 3 \pi/2$ for all
propagation directions in order for the medium to be left-handed.

SLPW is an example of such an anisotropic medium: it is strictly
left-handed (because $\pi/2 < \alpha < 3 \pi/2$ for all
propagation directions. Group and phase velocities exactly oppose
each other for four different propagation directions: $\vec{k}
\para \vec{e}_x, \vec{e}_y, (\vec{e}_x \pm \vec{e}_y)$. Due to the
anisotropy of the SLPW, radiation ''prefers'' to propagate along
one of the channels. Specifically, for $k_x > k_y$ radiation
preferentially flows along $x-$direction, and vice versa. Symmetry
requires that $\vec{P}
\para \vec{k}$ when $\vec{k} \para (\vec{e}_x + \vec{e}_y)$, so
there is a rapid transition near $\theta = \pi/4$ from the
preferential flow along $\vec{e}_x$ for $\theta < \pi/4$ to the
preferential flow along $\vec{e}_y$ for $\theta > \pi/4$. The
rapidity of the transition is determined by the corner curvature
of the structure. Note that the anisotropy of the SLPW is not
related to the periodicity of the structure. Numerical simulations
demonstrate that the SLPW is anisotropic even for small $|\vec{k}|
\ll d^{-1} \pi$ far from the edges of the Brillouin zone.

\subsection{Why is SLPW anisotropic for small wavenumbers}
The anisotropy of the SLPW for small $\vec{k}$ may be somewhat
surprising because it implies that the mode frequency is not an
analytic function of $\vec{k}$. Indeed, if the frequency is a
quadratic function of $\vec{k}$, i.~e.~$\omega^2 = \omega_c^2 +
g_{lm} k_l k_m$ (where $\omega_c$ is the cutoff frequency), then
the symmetry of the square lattice requires that $g_{kl} = D
\delta_{kl}$, where $D$ is the scalar and $\delta_{kl}$ is the
Kronecker delta. This would imply isotropy: from $\omega^2 =
\omega_c^2 + D \vec{k}^2$ follows that the group velocity
$\vec{v}_g = \partial \omega/\partial \vec{k} \parallel \vec{k}$
is parallel to the phase velocity. However, the frequency is not
an analytic function of $\vec{k}$. This is due to the fact that at
$\vec{k} = 0$ the mode is doubly-degenerate (see
Fig.~\ref{fig:left_right}). Below I illustrate using the
perturbation theory that the mode degeneracy implies anisotropy
for the SLPW.

The solutions of the Eq.~(\ref{eq:by1}) corresponding to the phase
shift $(\phi_x,\phi_y)$ per unit cell can be expressed as $H_z =
\tilde{H} \exp{i(k_x x + k_y y)}$, where $\tilde{H}$ is a periodic
function. Substituting $H_z$ into Eq.~(\ref{eq:by1}) yields
\begin{equation}
-\vec{\nabla} \cdot \left( \frac{\vec{\nabla} \tilde{H}}{\epsilon}
\right) + \frac{k^2 \tilde{H}}{\epsilon} - i \vec{k} \cdot \left(
\frac{2 \vec{\nabla} \tilde{H}}{\epsilon} + \tilde{H} \vec{\nabla}
\epsilon^{-1} \right) = \frac{\omega^2}{c^2} \tilde{H}.
\label{eq:by2}
\end{equation}
Equation (\ref{eq:by2}) can be recast as an eigenvalue problem
$(\hat{H}^{(0)} + \hat{V}^{(1)} + \hat{V}^{(2)}) \psi = \lambda
\psi$, where the distances are measured in units of $c/\omega_p$,
$\lambda = \omega^2/\omega_p^2$ is the eigenvalue, $\hat{H}^{(0)}
\psi = - \vec{\nabla} \cdot (\epsilon^{-1} \vec{\nabla} \psi)$ is
the unperturbed Hamiltonian, $\hat{V}^{(1)} \psi = - i \vec{k}
\cdot (2 \epsilon^{-1} \vec{\nabla} \psi + \psi \vec{\nabla}
\epsilon^{-1} )$ is the linear in $\vec{k}$ perturbation, and
$\hat{V}^{(2)} \psi = k^2 \psi/\epsilon$ is the quadratic in
$\vec{k}$ perturbation of the Hamiltonian. The total perturbation
$\hat{V} = \hat{V}_1 + \hat{V}_2$. The dielectric permittivity
$\epsilon$ the function of $\lambda$. However, for simplicity, I
will assume in the following that $\epsilon$ is frequency
independent. For definitiveness, assume that $\epsilon_0 = -0.4$
since, for $\vec{k} = 0$, $\epsilon_0 = 1 -
\omega_p^2/\omega_c^2$.

The unperturbed Hamiltonian has a discrete spectrum of
eigenfunctions $\psi_k$ and associated eigenvalues $\lambda_k$.
The lowest eigenvalue is $\lambda = 0$ which corresponds to
$\psi_0 = {\rm const}$. The next doubly-degenerate eigenvalue is
$\lambda_{I,II} \equiv \lambda_c = 0.71$. The corresponding
eigenfunctions are odd with respect to inversion:
$\psi_{I,II}(-\vec{x}) = - \psi_{I,II}(\vec{x})$. They can be
conveniently chosen such that $\psi_I(x,-y) = -\psi_I(x,y)$,
$\psi_I(-x,y) = \psi_I(x,y)$, and, correspondingly,
$\psi_{II}(x,-y) = \psi_{II}(x,y)$, $\psi_{II}(-x,y) =
-\psi_{II}(x,y)$. Note that $\psi_{I}$ and $\psi_{II}$ are
mutually orthogonal. The next closest to $\lambda_c$ eigenvalue is
$\lambda_3 = 1.4$. This a non-degenerate eigenvalue, with the
corresponding eigenfunction $\psi_3$ satisfying the following
symmetry properties: $\psi_3(-x,-y) = \psi_3(x,y)$, $\psi_3(x,-y)
= - \psi_3(x,y)$, and $\psi_3(-x,y) = - \psi_3(x,y)$.

The $\vec{k}$-dependent perturbation $V^{(1)}$ and $V^{(2)}$
removes the degeneracy between $\psi_I$ and $\psi_{II}$, resulting
in the left-handed and right-handed waves shown in
Fig.~\ref{fig:left_right} which have different frequencies as
indicated earlier. Calculating the mode frequencies in the limit
of small $\vec{k}$ can be done in the context of the perturbation
theory by solving the secular equation~\cite{sakurai}. Defining
the overlap integral $V_{ij}$ as
\[
V_{ij} = \int_0^d dx \int_0^d dy \ \psi_i^{\ast} \hat{V} \psi_j,
\]
where $\psi$'s are the unperturbed solutions, the secular equation
takes on the form which is familiar from quantum mechanics:
$V_{kl} c_l = \lambda_n \delta_{kl} c_l$. The summation over the
repeated index is assumed, and the sought after n-th perturbed
eigenfunction $\Psi_n$ is given by $\Psi_n = \sum_m c_m \psi_m$.
We are primarily interested in the behavior of the perturbed
solutions $\Psi_{-}$ (lower-frequency left-handed mode) and
$\Psi_{+}$ (upper-frequency right-handed mode) which, in the limit
of $\vec{k} = 0$, become $\psi_{I}$ and $\psi_{II}$.

Strictly, all unperturbed eigenfunctions of $\hat{H}_0$ have to be
included in the summation. For simplicity, I only include the two
degenerate eigenfunctions $\psi_{I}$, $\psi_{II}$, and the
eigenfunctions $\psi_0$ and $\psi_3$ which have the closest
eigenvalues $\lambda = 0$ and $\lambda = \lambda_3$. In this
truncated model, the degenerate doublet of unperturbed
eigenfunctions ($\psi_I$,$\psi_{II}$) is ``surrounded'' by the two
singlets $\psi_0$ and $\psi_3$. The four eigenfunctions
($\psi_0,\psi_{I},\psi_{II},\psi_3$) are taken to be real and
normalized to unity.

Simple symmetry arguments can be used to demonstrate that
$V_{I,II}= 0$ and $V_{I,I} = V_{II,II} = V_{I,I}^{(2)} \equiv A
k^2$, where $A = (\epsilon^{-1})_{I,I}$ is a constant. Thus, the
perturbation $\hat{V}$ does not mix $\psi_{I}$ and $\psi_{II}$,
and does not remove the degeneracy of the doublet
($\psi_I$,$\psi_{II}$). It is the mixing between the doublet and
the two singlets via $\hat{V}_1$ that lifts the degeneracy, and
introduces the anisotropy of the perturbed solutions $\Psi_{-}$
and $\Psi_{+}$. The relevant non-vanishing matrix elements are
$V_{I,0}^{(1)} = V_{10}$, $V_{II,0}^{(1)} = V_{20}$,
$V_{I,3}^{(1)} = V_{13}$, and $V_{II,3}^{(1)} = V_{23}$. It can be
shown that the effect of the non-vanishing matrix elements $V_{00}
\equiv B k^2$ and $V_{33} \equiv C k^2$ (where $B$ and $C$ are
constants) on the doublet is of order $|\vec{k}|^4$. Hence,
$V_{00}$ and $V_{33}$ are neglected. Also by symmetry, $V_{03} =
0$.

The corresponding secular equation for computing the perturbed
frequencies of ($\Psi_{+}$,$\Psi_{-}$) is obtained by solving the
following equation~\cite{sakurai} for the eigenvalue shift $\delta
\lambda$:
\begin{equation}
\left( \begin{array}{cccc}
  0 & 0 & V_{10} & V_{13} \\
  0 & 0 & V_{20} & V_{23}  \\
  V_{10}^{\ast} & V_{20}^{\ast} & (-\lambda_c) & 0 \\
  V_{13}^{\ast} & V_{23}^{\ast} & 0 & (\lambda_3 - \lambda_c) \\
\end{array} \right)
  \left(  \begin{array}{c}
  c_{I} \\
  c_{II} \\
  c_{0} \\
  c_{3} \\
\end{array} \right) = \bar{\lambda} \left(  \begin{array}{c}
  c_{I} \\
  c_{II} \\
  c_{0} \\
  c_{3} \\
\end{array} \right),
\label{eq:secular1}
\end{equation}
where $\bar{\lambda} \equiv (\delta \lambda) - Ak^2$. In the
vicinity of $\bar{\lambda} = 0$, the quatric in $\bar{\lambda}$
Eq.~(\ref{eq:secular1}) becomes quadratic:
\begin{equation}
\bar{\lambda}^2 + \bar{\lambda} \left( \frac{|V_{10}|^2 +
|V_{20}|^2}{E_0} + \frac{|V_{13}|^2 + |V_{23}|^2}{E_3} \right) +
\frac{|\det{Q}|^2}{E_0 E_3} = 0, \label{eq:secular2}
\end{equation}
where $E_0 = -\lambda_c$, $E_3 = \lambda_3 - \lambda_c$, and $Q$
is the interaction matrix:
\begin{equation}
Q = \left(
\begin{array}{cc}
V_{10} & V_{13} \\
V_{20} & V_{23} \\
\end{array}
\right). \label{eq:Q_matrix}
\end{equation}

Again, symmetry considerations completely determine the functional
dependence of the matrix elements on $\vec{k}$. For example, it
can be shown that for the SLPW $V_{10} = iM k_y$, $V_{20} =
iMk_x$, $V_{13} = i N k_x$, and $V_{23} = i N k_y$. Therefore, the
bracketed term in Eq.~(\ref{eq:secular1}) is the function of
$|\vec{k}|^2$ only. The vanishing of $\det{Q}$ would then
guarantee that $\bar{\lambda}$ is an analytic function of
$\vec{k}$. Not surprisingly, $(\delta \lambda) = (\omega^2 -
\omega_c^2)/\omega_p^2$ would be an isotropic function of
$|\vec{k}|^2$. For the SLPW, however, $|\det{Q}|^2 = M^2 N^2
(k_x^2 - k_y^2)^2$ does not vanish, and
\begin{equation}
\bar{\lambda} =  -\frac{|V_{10}|^2 - |V_{20}|^2}{2E_0} +
\frac{|V_{13}|^2 + |V_{23}|^2}{2E_3} \pm \sqrt{\left(
\frac{|V_{10}|^2 + |V_{20}|^2}{2E_0} + \frac{|V_{13}|^2 +
|V_{23}|^2}{2E_3} \right)^2 - \frac{|\det{Q}|^2}{E_0 E_3}},
\label{eq:singular_sol}
\end{equation}
where the $+$ ($-$) signs correspond to $\Psi_{+}$ (right-handed)
and $\Psi_{-}$ (left-handed) modes, respectively.

Note from Eq.~(\ref{eq:singular_sol}) that $\bar{\lambda}$ is not
an analytic function. It is also manifestly anisotropic due to the
$\vec{k}$-dependence of $Q$. The two properties (anisotropy and
the non-analyticity) go hand in hand: analytic dependence on
$\omega$ on $\vec{k}$ implies isotropy in the limit of small
$\vec{k}$. In Section~\ref{sec:tlpw} it is found that a periodic
lattice with higher symmetry (Triangular Lattice Photonic
Waveguide) is isotropic. This may be explained by the vanishing of
$\det{Q}$ due to the higher degree of symmetry of the TLPW.

\section{Isotropy of Triangular Lattice Photonic Waveguide}
\label{sec:tlpw}

\begin{figure}[h]
\centering {\hspace{-0pt}
\epsfig{file=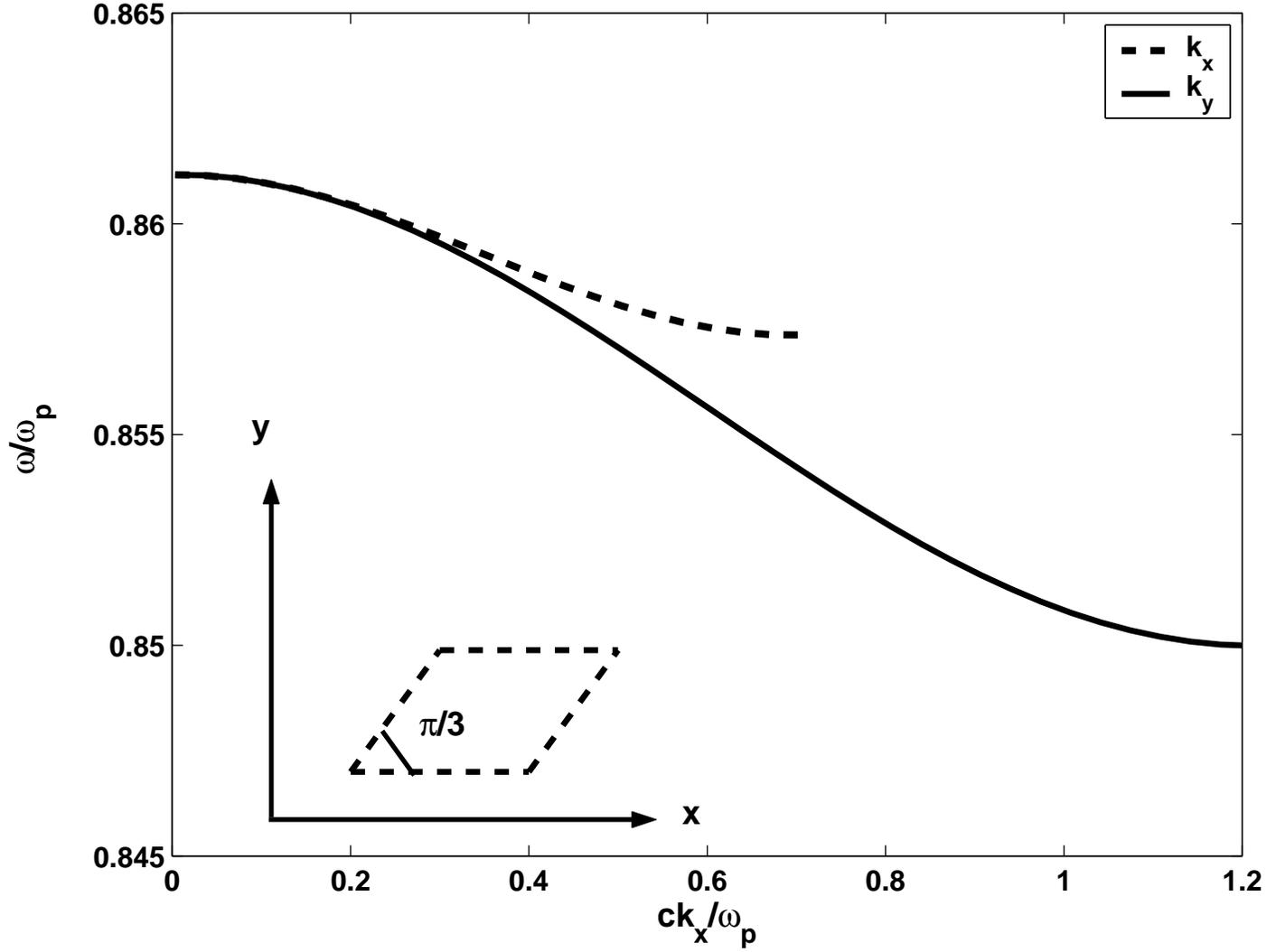,angle=-0}}
\vspace{1pt}
\caption{Dispersion relation $\omega$ v.~s.
$|\vec{k}|$ for a triangular arrangement of dielectrics as in
Fig.~\ref{fig:figure_structures}(d). Equilateral parallelogram
with $d = 3 c/\omega_p$ and $\beta = \pi/3$ opening angle --
elementary cell of the photonic structure. Solid line: $\vec{k} =
k \vec{e}_y$, $0 < kd < 2 \pi/\sqrt{3}$; dashed line: $\vec{k} = k
\vec{e}_x$, $0 < kd < 2 \pi/3$. Channel widths $2b = 0.6
c/\omega_p$, dielectric edges smoothed with radius $r_b = b$.}
\label{fig:triang_band}
\end{figure}

Photonic waveguide can be made much more isotropic if dielectrics
are arranged in a triangular lattice, as shown in
Fig.~\ref{fig:figure_structures}(d), forming a Triangular Lattice
Photonic Waveguide (TLPW). This arrangement has a higher degree of
symmetry than SLPW, and results in a composite medium which is
perfectly isotropic for small $\vec{k}$ well inside the Brillouin
zone. The elementary cell of the TLPW and the dispersion relation
$\omega$ v.~s.~$\vec{k}$ for two orthogonal directions of $\vec{k}
= \vec{e}_{x,y} k$ are shown in Fig.~\ref{fig:triang_band}. The
two dispersion curves, which are identical for $|\vec{k}| \ll
\pi/d$ (establishing the isotropy), are drawn to the respective
edges of the Brillouin zone: $0 < k_x d < 2\pi/3$ and $0 < k_y d <
2\pi/\sqrt{3}$~\cite{smirnova_jap}.

\begin{figure}[h]
\centering {\hspace{-0pt} \epsfig{file=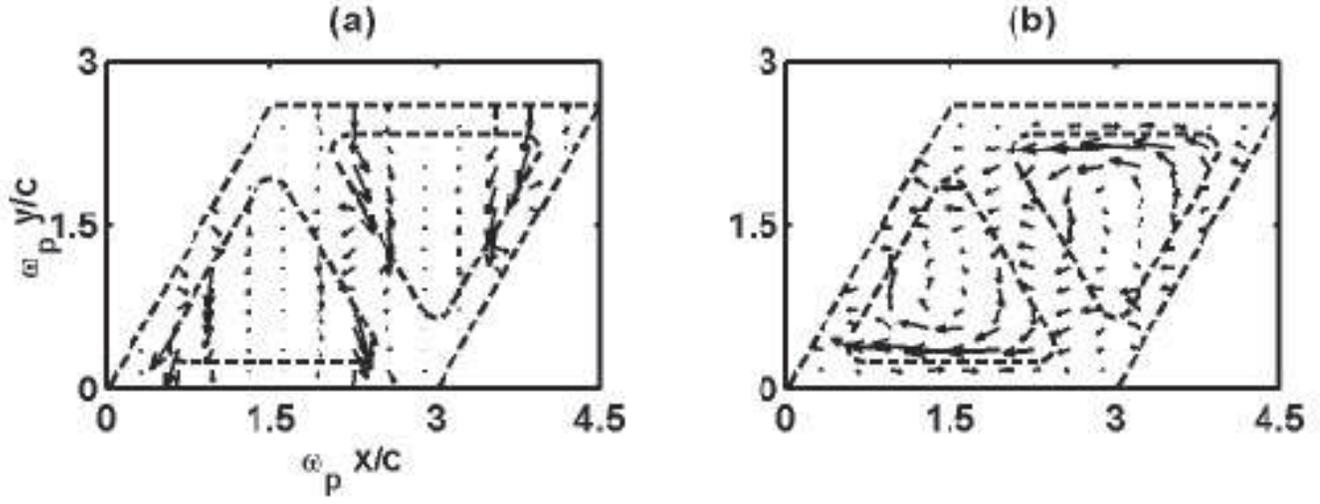,angle=-0}}
\vspace{1pt} \caption{Power flow in a Triangular Lattice Photonic
Waveguide. Structure parameters same as in
Fig.~\ref{fig:triang_band}. Wavenumber $k = d^{-1} \pi/6$, and (a)
$\vec{k} = k \vec{e}_y$ (no vacuum/cladding interfaces parallel to
$\vec{k}$); (b) $\vec{k} = k \vec{e}_x$, (interface along
$\vec{k}$) } \label{fig:flow_kxky}
\end{figure}

These two directions are chosen because they exhibit the
maximum possible anisotropy. Indeed, for $\vec{k} = k \vec{e}_x$
there exists a vacuum/cladding interface along which the
electromagnetic energy can flow in $x-$direction,
Fig.~\ref{fig:flow_kxky}(b). No such interface exists for $\vec{k}
= k \vec{e}_y$, Fig.~\ref{fig:flow_kxky}(a). Thus, the local
Poynting flux patterns are very different for these two
directions, as can be seen by comparing
Figs.~\ref{fig:flow_kxky}(a,b).
For the parameters of Fig.~(\ref{fig:flow_kxky}), $k =
d^{-1} \pi/6$, the index of refraction is $n = -ck/\omega = -0.2$.
Index of refraction can be readily tuned in both directions by
adjusting the parameters of the TLPW: periodicity $d$ and channel
width $2b$.

Despite the differences in the flow patterns, the cell-integrated
fluxes $\vec{P}_{\rm av}$ are identical for all directions of $\vec{k}$,
and so are the frequencies: $\omega = 0.86 \omega_p$. Numerical results
unambiguously confirm that, for small $|\vec{k}|$, phase and group
velocities exactly oppose each other. At least for this particular
LHM, the claim of Valanju~{\it et.~al.~} \cite{valanju} that
$\alpha = \angle (\vec{v}_{\rm ph}, \vec{v}_{\rm g}) \neq \pi$ is
not confirmed.  Anisotropy for large $|\vec{k}|$ is merely the
consequence of the periodicity of the photonic structure.

The angle $\alpha = \angle (\vec{k},\vec{P}_{\rm av})$ between the
phase and group velocities was calculated as the function of the
propagation angle $\theta = \cos^{-1}(k_x/k)$. The propagation
wavenumber of magnitude $|\vec{k}| = d^{-1}\pi/4$ was rotated by
varying $0 < \theta < 2\pi/3$. The angle $\alpha$ for the TLPW was
found to vary between $0.994 \pi < \alpha < 1.006 \pi$. This
stands in sharp contrast with the significant deviation of
$\alpha$ from $\pi$ for the SLPW as shown in
Fig.~(\ref{fig:poynt_angle}). Therefore, it is concluded that,
unlike the SLPW, the TLPW is an isotropic (uniaxial) left-handed
structure. I speculate that the isotropy is related to the fact
that $\det{Q}$ given by Eq.~(\ref{eq:Q_matrix}) vanishes for TLPW
but not for the SLPW.

\section{Conclusions}
\label{sec:concl}

In conclusion, several photonic structures supporting left-handed
waves were considered. The structures consist of dielectrics with
the dielectric permittivities $\epsilon$ of the opposing signs,
enabling the propagation of the surface waves. Two types of
surface waves, plasmons and phonons, were considered. A specific
material (SiC) was suggested as the promising candidate for making
left-handed photonic structures, as well as for developing a
``perfect'' lens for enhanced near-field imaging.

The most promising two-dimensional structure with negative index
of refraction appears to be the Triangular Lattice Photonic
Waveguide because it is isotropic (uniaxial) in the wave
propagation plane. The simpler left-handed structure, the Square
Lattice Photonic Waveguide, was shown to be anisotropic even for
infinitesimally small propagation wavenumbers. This was shown to
be the consequence of the non-analytic dependence of the wave
frequency on its wavenumber.

I would like to acknowledge the enlightening conversations with
Dr.~D.~J.~Bergman and Dr.~R.~C.~McPhedran on the anisotropy of the
square lattice photonic waveguide.

\end{document}